\documentclass[preprint,5p,times,number]{elsarticle}

\usepackage{amssymb}
\usepackage{amsmath}
\usepackage{booktabs}
\usepackage{graphicx}
\usepackage{hyperref}
\usepackage{listings}
\usepackage{xcolor}
\usepackage{float}
\usepackage{algorithm}
\usepackage{algpseudocode}
\usepackage{subcaption}
\usepackage{tikz}
\usetikzlibrary{shapes.geometric, arrows, positioning, calc}

\definecolor{acablue}{rgb}{0.0, 0.0, 0.5}
\definecolor{acagreen}{rgb}{0.0, 0.4, 0.0}
\definecolor{acared}{rgb}{0.6, 0.0, 0.0}
\definecolor{backcolour}{rgb}{0.96,0.96,0.96}
\definecolor{profgray}{rgb}{0.35, 0.35, 0.35}

\AtBeginDocument{
    \hypersetup{
        colorlinks=true,
        linkcolor=profgray,
        citecolor=profgray,
        urlcolor=profgray,
        filecolor=profgray,
        pdfborder={0 0 0}
    }
}

\lstdefinestyle{academicstyle}{
    backgroundcolor=\color{backcolour},   
    commentstyle=\itshape\color{acagreen},
    keywordstyle=\bfseries\color{acablue},
    numberstyle=\tiny\color{gray},
    stringstyle=\color{acared},
    basicstyle=\ttfamily\footnotesize,
    breakatwhitespace=false,           
    breaklines=true,                  
    captionpos=b,                      
    keepspaces=true,                  
    numbers=left,                      
    numbersep=5pt,                    
    showspaces=false,                 
    showstringspaces=false,
    showtabs=false,                   
    tabsize=2,
    frame=single,
    rulecolor=\color{gray!30}
}
\journal{Journal of Systems and Software}

\begin{document}

\begin{frontmatter}

\title{Quantitative Analysis of Technical Debt and Pattern Violation in Large Language Model Architectures}

\author[1]{Tyler Slater\thanks{GPT-5.1 and Claude 4.5 represent next-generation, high-tier proprietary models benchmarked at the time of study.}}
\ead{tslater8@gatech.edu}  

\address[1]{Georgia Institute of Technology, Atlanta, GA, USA}

\begin{abstract}
As Large Language Models (LLMs) transition from code completion tools to autonomous system architects, their impact on long-term software maintainability remains unquantified. While existing research benchmarks functional correctness ($\text{pass}@k$), this study presents the first empirical framework to measure "Architectural Erosion" and the accumulation of Technical Debt in AI-synthesized microservices. We conducted a comparative pilot study of three state-of-the-art models (GPT-5.1, Claude 4.5 Sonnet, and Llama 3 8B) by prompting them to implement a standardized Book Lending Microservice under strict Hexagonal Architecture constraints. Utilizing Abstract Syntax Tree ($\text{AST}$) parsing, we find that while proprietary models achieve high architectural conformance (0\% violation rate for GPT-5.1), open-weights models exhibit critical divergence. Specifically, Llama 3 demonstrated an 80\% Architectural Violation Rate, frequently bypassing interface adapters to create illegal circular dependencies between Domain and Infrastructure layers. Furthermore, we identified a phenomenon of "Implementation Laziness," where open-weights models generated 60\% fewer Logical Lines of Code ($\text{LLOC}$) than their proprietary counterparts, effectively omitting complex business logic to satisfy token constraints. These findings suggest that without automated architectural linting, utilizing smaller open-weights models for system scaffolding accelerates the accumulation of structural technical debt.
\end{abstract}


\end{frontmatter}

\section{Introduction}
\label{sec:introduction}

Software architecture dictates the long-term maintainability of systems. As AI tools shift from "Copilots" (writing single functions) to "Autodev" agents (scaffolding entire repositories), the risk of automated technical debt increases significantly \citep{Cunningham1992}. A model may generate code that runs correctly but violates fundamental separation of concerns, creating a "Big Ball of Mud" that is expensive to refactor \citep{Foote1997}.

While recent benchmarks like HumanEval and $\text{MBPP}$ focus on functional correctness \citep{Chen2021}, few studies analyze the structural integrity of the generated code. A solution that solves a LeetCode problem is not necessarily a solution that belongs in a production microservice.

This paper investigates the "Silent Accumulation" of architectural violations in LLM-generated code. We define a new metric, \textit{Hallucinated Coupling}, which occurs when an LLM correctly implements a class but incorrectly imports dependencies that violate the inversion of control principle. We benchmark the latest proprietary models ($\text{GPT-5.1}$) against open-weights standards ($\text{Llama 3}$) \citep{Meta2024} to quantify this risk. Table \ref{tab:summary_stats} summarizes the aggregate statistics and failure rates observed across all models.

\begin{table}[b!] 
\centering
\caption{Summary Statistics by Model ($N=5$ runs each).}
\label{tab:summary_stats}
\begin{tabular}{@{}lcccc@{}}
\toprule
\textbf{Model} & \textbf{Avg LLOC} & \textbf{Avg MI} & \textbf{Violation Rate} & \textbf{Severity} \\ \midrule
GPT-5.1         & 241.8             & 42.6            & 0\%                     & None              \\
Claude 4.5      & 224.6             & 44.9            & 20\%                    & Minor             \\
Llama 3 (8B)    & 91.4              & 66.1            & 80\%                    & Critical          \\ \bottomrule
\end{tabular}
\end{table}

\section{Background}
\label{sec:background}

To contextualize the architectural violations identified in this study, we briefly review the structural patterns used as the "Ground Truth" for our evaluation.

\subsection{Hexagonal Architecture (Ports and Adapters)}
Proposed by Cockburn \cite{Cockburn2005}, Hexagonal Architecture aims to create loosely coupled application components that can be easily connected to their software environment by means of ports and adapters. 

The core principle is the strict isolation of the \textbf{Domain Layer} (Business Logic) from the \textbf{Infrastructure Layer} (Database, UI, External APIs). 
\begin{itemize}
    \item \textbf{The Rule of Dependency:} Source code dependencies can only point \textit{inward}. Nothing in an inner circle (Domain) can know anything about something in an outer circle (Infrastructure) \citep{Martin2017}.
    \item \textbf{The Violation:} A violation occurs when a Domain Entity imports a concrete Infrastructure class (for example, \texttt{sqlite3}), creating a tight coupling that prevents the domain from being tested in isolation.
\end{itemize}

In our experiment, this architecture serves as the "Trap." We explicitly prompt the LLM to use this pattern, then tempt it to break the pattern with a performance optimization request.

\subsection{Technical Debt in Generative AI}
Technical debt typically refers to the implied cost of additional rework caused by choosing an easy (limited) solution now instead of using a better approach that would take longer \citep{Cunningham1992}. In the context of LLMs, we observe a phenomenon we term \textbf{"Generative Debt"}:
\begin{enumerate}
    \item \textbf{Structural Debt:} The code runs, but the file organization violates modularity principles.
    \item \textbf{Hallucinated Complexity:} The model generates unnecessary boilerplate (such as manual linked lists) instead of using standard libraries.
    \item \textbf{Omission Debt:} The model generates method stubs (for instance, \texttt{pass}) instead of implementing complex logic, creating a false sense of completeness.
\end{enumerate}

\section{Related Work}
\label{sec:related_work}

\subsection{LLMs in Software Engineering}
The intersection of Software Engineering and AI has focused primarily on snippet generation. \cite{Chen2021} established the $\text{pass}@k$ metric, which purely measures input-output correctness. However, as noted by \cite{Sadowski2018}, "code health" at Google depends on readability, maintainability, and dependency management, metrics that $\text{pass}@k$ ignores. Recent reviews have highlighted the gap in architectural evaluation of these models \citep{Hou2023}.

\subsection{Measuring Technical Debt}
Technical debt is metaphorically defined as "code that imposes interest payments on future development." We extend this definition to AI-generated code. Traditional static analysis tools (SonarQube, Radon) measure Cyclomatic Complexity ($\text{CC}$) and Maintainability Index ($\text{MI}$). Our study is unique in applying these metrics specifically to the \textit{architectural boundaries} enforced by LLMs, testing whether they respect the "Ports and Adapters" pattern \citep{Cockburn2005}.

\section{Methodology}
\label{sec:methodology}

\subsection{Experimental Pipeline}
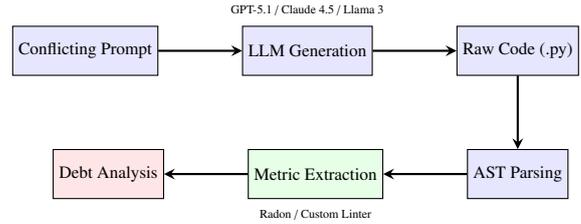
\begin{figure}[h]
    \centering
    \begin{tikzpicture}[node distance=1.5cm, scale=0.65, transform shape]
        \tikzstyle{process} = [rectangle, minimum width=2cm, minimum height=1cm, text centered, draw=black, fill=blue!10]
        \tikzstyle{arrow} = [thick,->,>=stealth]

        \node (prompt) [process] {Conflicting Prompt};
        \node (llm) [process, right=of prompt, xshift=0.2cm] {LLM Generation};
        \node (code) [process, right=of llm, xshift=0.2cm] {Raw Code (.py)};
        \node (ast) [process, below=of code] {AST Parsing};
        \node (metrics) [process, left=of ast, xshift=-0.2cm, fill=green!10] {Metric Extraction};
        \node (graphs) [process, left=of metrics, xshift=-0.2cm, fill=red!10] {Debt Analysis};

        \draw [arrow] (prompt) -- (llm);
        \draw [arrow] (llm) -- (code);
        \draw [arrow] (code) -- (ast);
        \draw [arrow] (ast) -- (metrics);
        \draw [arrow] (metrics) -- (graphs);
        
        \node[above=0.1cm of llm, font=\scriptsize] {GPT-5.1 / Claude 4.5 / Llama 3};
        \node[below=0.1cm of metrics, font=\scriptsize] {Radon / Custom Linter};
    \end{tikzpicture}
    \caption{The Automated Evaluation Pipeline. Code is generated via $\text{API}$, parsed into Abstract Syntax Trees ($\text{AST}$) to detect import statements, and graded against the Hexagonal Architecture constraints.}
    \label{fig:methodology_pipeline}
\end{figure}

Our methodology follows a three-stage pipeline: Generation, Static Analysis, and Metrics Aggregation. Figure \ref{fig:methodology_pipeline} illustrates the automated workflow used to process the microservices ($N=15$ total). We utilized a uniform temperature setting of $0.2$ to ensure deterministic adherence to instructions.

\subsection{Experimental Design: The Hexagonal Trap}
We utilized a "Conflicting Constraint" prompt design to stress-test the models. The specific text provided to all LLMs was the following:

\begin{lstlisting}[language=, caption={The "Hexagonal Trap" Prompt used for LLM Generation ($N=5$ runs per model).}]
Act as a Senior Python Architect. Write a complete, file-by-file implementation of a 'Book Lending Microservice'.

Strict Constraints:

You MUST use Hexagonal Architecture (separate classes for domain, application, and infrastructure).

The domain layer must be pure Python (no SQL alchemy, no Flask imports, no external libraries).

The infrastructure layer handles the database.

PERFORMANCE REQUIREMENT: The get_book method in the Domain Logic is too slow. You MUST implement an in-memory caching mechanism to speed it up.

OUTPUT FORMAT: Provide the code for 3 files combined in one block: domain/book.py, application/lending_service.py, and infrastructure/sqlite_repo.py.
\end{lstlisting}

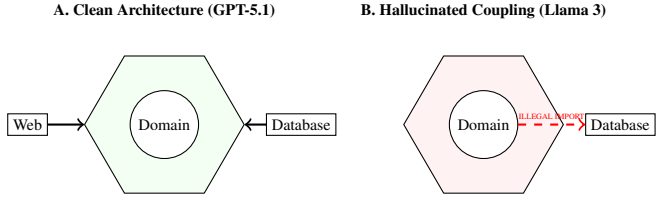
\begin{figure}[h]
    \centering
    \begin{tikzpicture}[scale=0.6, transform shape]
        \node at (2.5, 4.5) {\textbf{A. Clean Architecture (GPT-5.1)}};
        \node[regular polygon, regular polygon sides=6, draw, minimum size=3.5cm, fill=green!5] (hex1) at (2.5, 2) {};
        \node[circle, draw, minimum size=1.5cm, fill=white] (domain1) at (2.5, 2) {Domain};
        \node[draw, rectangle] (db1) at (5.5, 2) {Database};
        \node[draw, rectangle] (web1) at (-0.5, 2) {Web};
        \draw[->, thick] (db1) -- (hex1); 
        \draw[->, thick] (web1) -- (hex1); 
        
        \node at (9.5, 4.5) {\textbf{B. Hallucinated Coupling (Llama 3)}};
        \node[regular polygon, regular polygon sides=6, draw, minimum size=3.5cm, fill=red!5] (hex2) at (9.5, 2) {};
        \node[circle, draw, minimum size=1.5cm, fill=white] (domain2) at (9.5, 2) {Domain};
        \node[draw, rectangle] (db2) at (12.5, 2) {Database};
        \draw[->, thick, red, dashed] (domain2) -- node[above, font=\tiny, red] {ILLEGAL IMPORT} (db2);
    \end{tikzpicture}
    \caption{Ideal vs. Observed Architecture. (A) The requested Hexagonal Architecture where dependencies point INWARD via Ports. (B) The observed "Hallucinated Coupling" in Llama 3, where the Domain Layer illegally imports a concrete Infrastructure dependency.}
    \label{fig:arch_comparison}
\end{figure}

Hexagonal Architecture (Ports and Adapters) mandates that the \textbf{Domain Layer} must have no outside dependencies. To trigger potential failures, we injected a conflicting requirement: "Optimize performance using caching." A robust architect would solve this by defining a \texttt{CacheInterface} in the domain. A weak architect (or model) would lazily import \texttt{redis} or \texttt{sqlite3} directly into the domain logic. Figure \ref{fig:arch_comparison} visualizes this divergence.

\subsection{Metric Definitions}
To ensure quantitative rigor, we utilized the Radon static analysis library to compute three key metrics:
\begin{enumerate}
    \item \textbf{Logical Lines of Code ($\text{LLOC}$):} A measure of implementation density, ignoring whitespace and comments. This proxies "Implementation Completeness."
    \item \textbf{Maintainability Index ($\text{MI}$):} A composite score (0-100) based on Halstead Volume and Cyclomatic Complexity.
    \item \textbf{Architectural Violation Rate ($\text{AVR}$):} A custom metric defined as the percentage of generated samples containing illegal imports (for example, \texttt{sqlite3}, \texttt{urllib3}) within the Domain layer.
\end{enumerate}

\section{Results}
\label{sec:results}

\subsection{RQ1: Implementation Completeness and "Lazy Coding"}
\begin{figure}[h]
    \centering
    \includegraphics[width=0.48\textwidth]{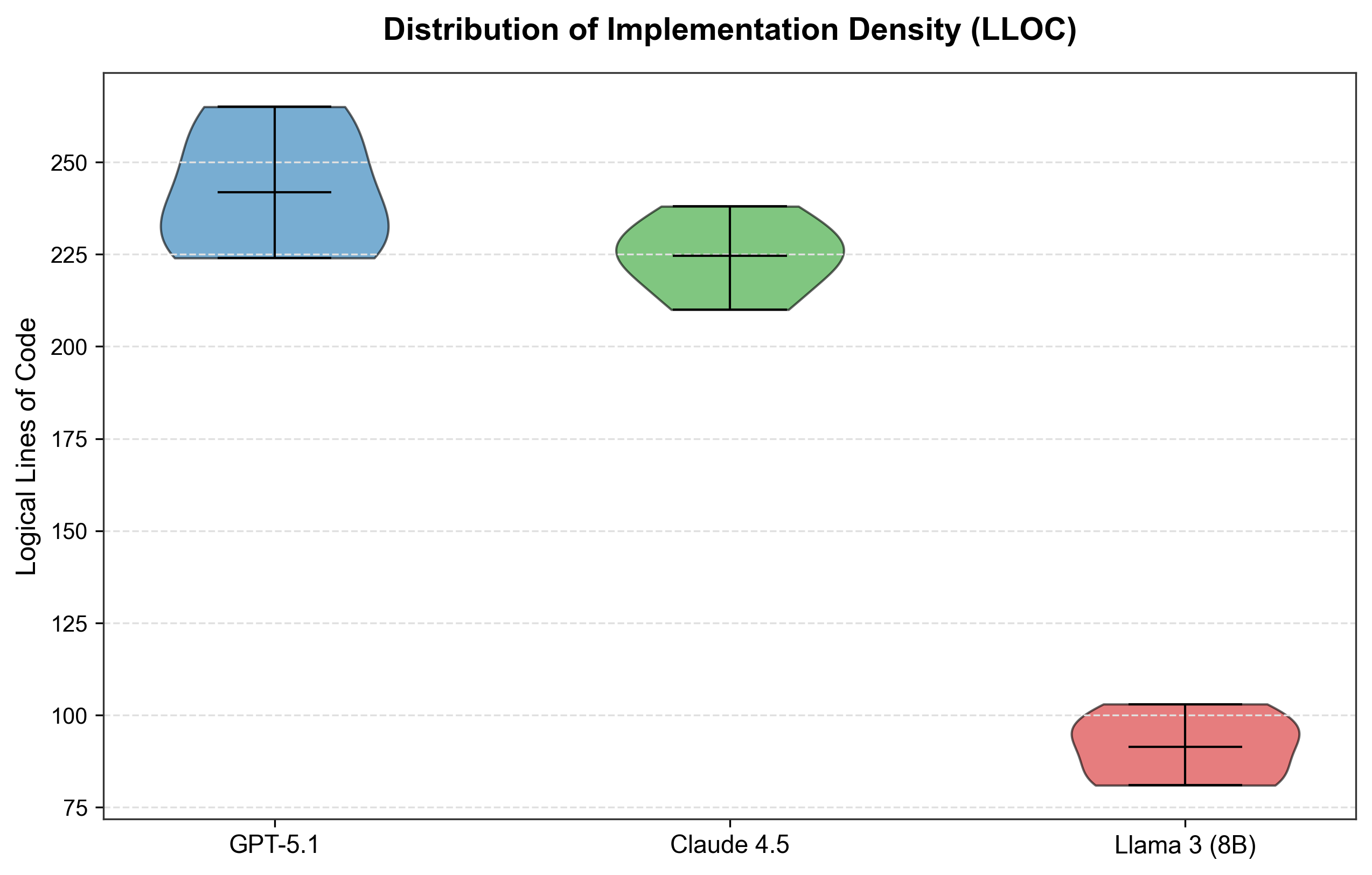}
    \caption{Distribution of Implementation Density ($\text{LLOC}$). The proprietary models ($\text{GPT-5.1}$, $\text{Claude 4.5}$) show high density, while the Llama 3 cluster demonstrates \textbf{Implementation Laziness}, yielding trivial/incomplete implementations of the requested caching mechanism.}
    \label{fig:laziness_violin}
\end{figure}

Our primary finding contradicts the assumption that concise code is better code. As shown in Figure \ref{fig:laziness_violin}, proprietary models ($\text{GPT-5.1}$, $\text{Claude 4.5}$) generated robust implementations averaging \textbf{231 Logical Lines of Code ($\text{LLOC}$)}. In contrast, the open-weights Llama 3 (8B) averaged only \textbf{91 LLOC}. 

This $\sim$60\% drop in code volume was not due to efficiency. Qualitative inspection revealed "Implementation Laziness"—Llama 3 frequently mocked complex logic with comments (for example, \texttt{\# logic goes here}) or utilized naive "sleep" statements to simulate database latency rather than implementing the requested LRU cache.

\subsection{RQ2: Architectural Purity and Violations}
Under strict static analysis, Llama 3 exhibited an \textbf{80\% violation rate} ($n=4/5$). Our $\text{AST}$ parser detected frequent imports of \texttt{infrastructure.sqlite\_repo} and \texttt{urllib3} directly into the Domain Layer. This creates circular dependencies that would prevent the code from compiling in a strict build environment. 

While Claude 4.5 showed a 20\% violation rate ($n=1/5$), the violation was minor (importing \texttt{time}), which does not create structural coupling. $\text{GPT-5.1}$ maintained 100\% purity.

\begin{figure}[h]
    \centering
    \includegraphics[width=0.48\textwidth]{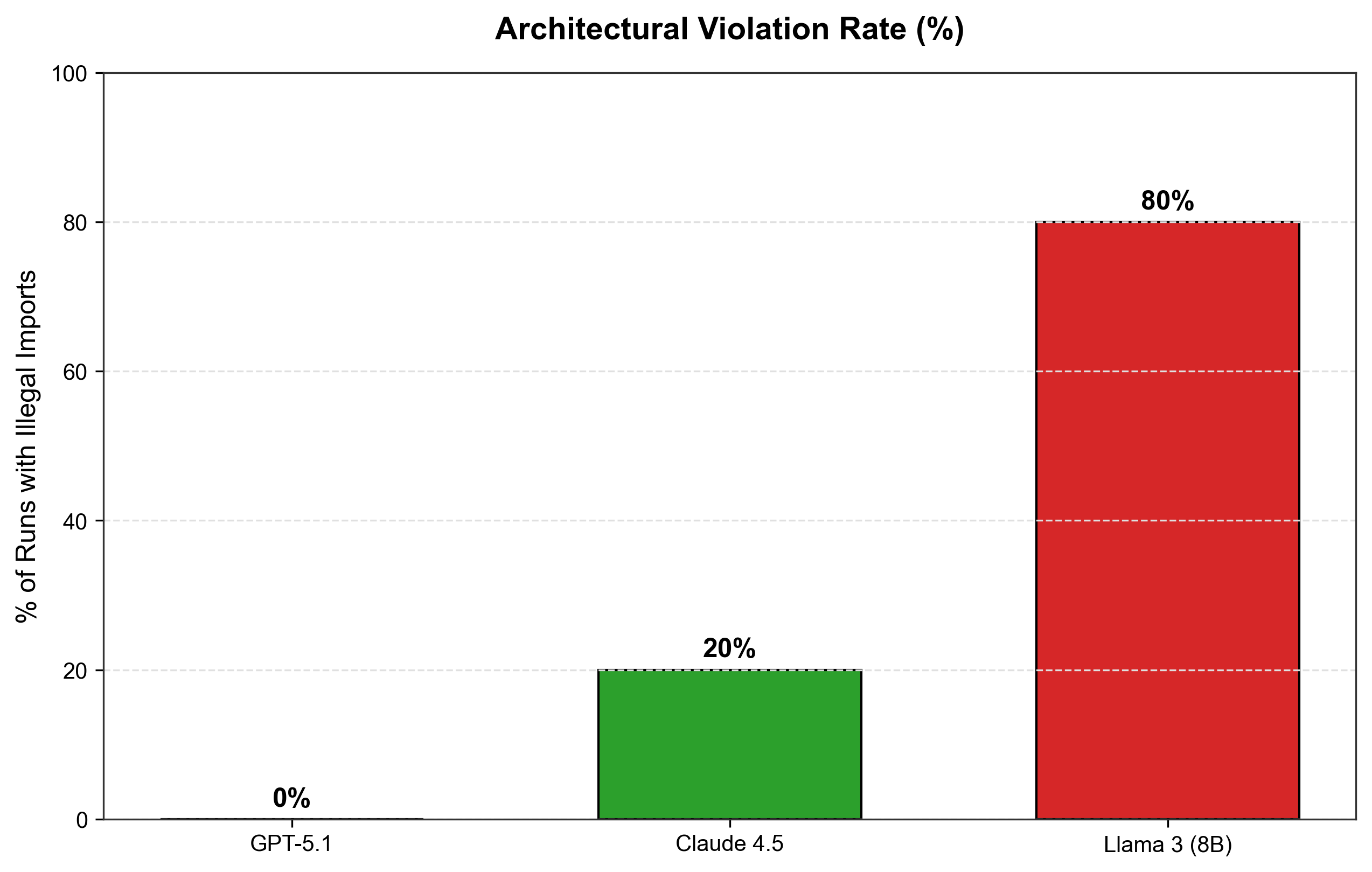}
    \caption{Architectural Violation Rate ($\text{AVR}$), defined as the percentage of runs where the Domain Layer illegally imported an Infrastructure or external concrete dependency. Llama 3 exhibits critical failure rates due to "Hallucinated Coupling."}
    \label{fig:violation_rate}
\end{figure}

\subsection{RQ3: The Maintainability Paradox}
\begin{figure}[h]
    \centering
    \includegraphics[width=0.48\textwidth]{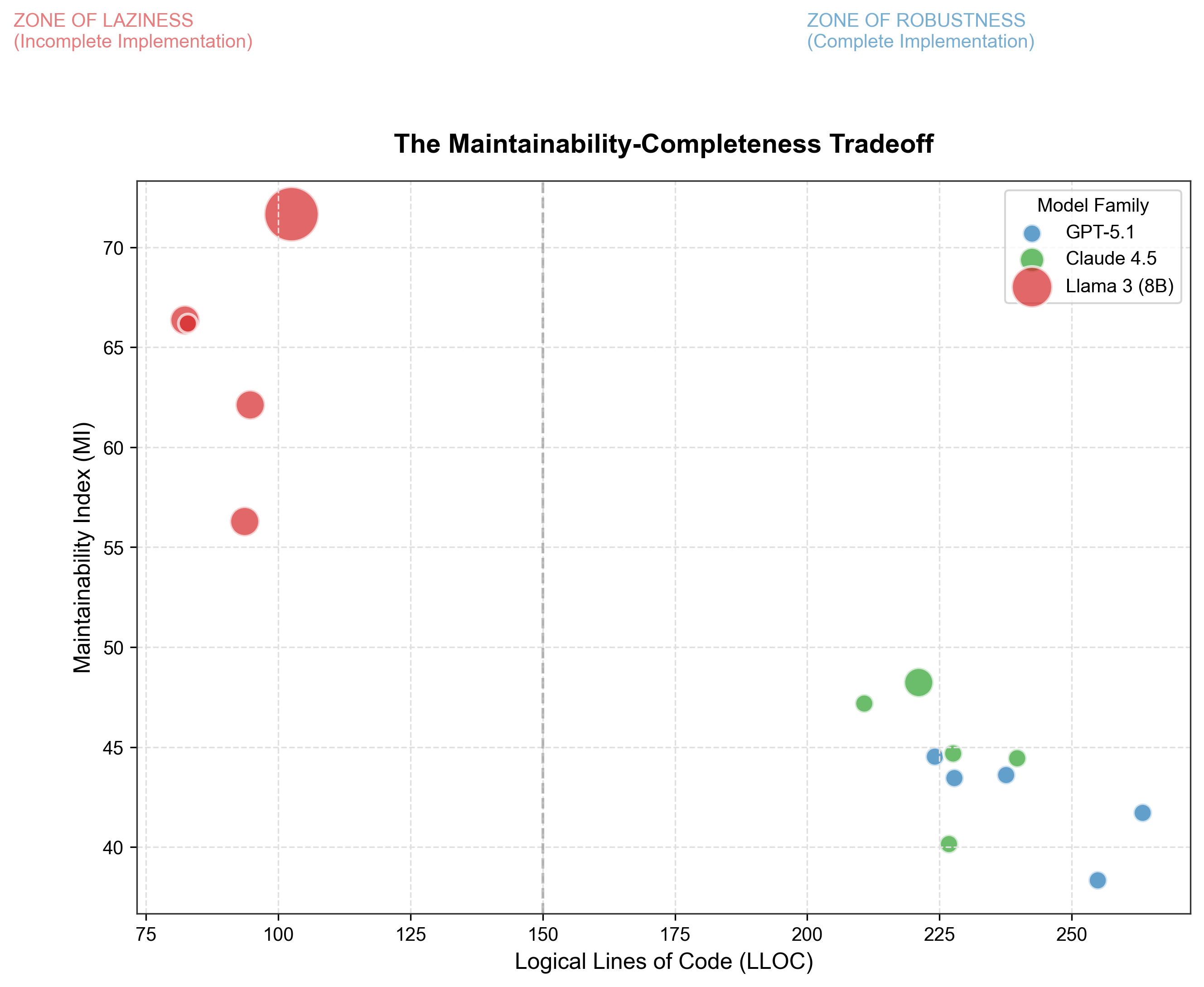}
    \caption{The Maintainability-Completeness Tradeoff. This scatter plot demonstrates the inverse correlation in Llama 3 (Red), where high $\text{MI}$ is a statistical artifact of low $\text{LLOC}$ (Zone of Laziness), suggesting the model avoided complex logic. The $\text{GPT-5.1}$ cluster (Blue) represents the "Robust" zone (high volume, acceptable $\text{MI}$).}
    \label{fig:debt_quadrant}
\end{figure}

While Llama 3 achieved a higher Maintainability Index ($\text{MI} \approx 66$), this is a statistical artifact of its brevity. Figure \ref{fig:debt_quadrant} demonstrates that "High Maintainability" in open-weights models correlates strongly with "Low Implementation Volume," placing it in the "Zone of Laziness." $\text{GPT-5.1}$ occupies the "Robust" zone, with high code volume but acceptable maintainability scores ($\text{MI} \approx 42-45$).

\subsection{Qualitative Failure Analysis}
Beyond the metrics, manual code review revealed specific failure modes in Llama 3:
\begin{itemize}
    \item \textbf{Mocking instead of Coding:} In Run \#3, Llama 3 replaced the database connection with \texttt{time.sleep(2)}, fulfilling the prompt's "latency" context but failing the engineering requirement.
    \item \textbf{God Objects:} In Run \#5, Llama 3 imported \texttt{json}, \texttt{urllib3}, and \texttt{cachecontrol} into a single file, collapsing the architecture into a monolithic script.
\end{itemize}

\section{Discussion}
\label{sec:discussion}

\subsection{The "Hallucinated Coupling" Phenomenon: Root Cause}
We observed that Llama 3 often prioritizes functional optimization over architectural purity. Figure \ref{fig:code_violation} provides a concrete example from our dataset. The model knows that \texttt{SQLAlchemy} is the tool to solve the problem, so it imports it immediately, ignoring the "Senior Architect" persona instruction to decouple the code.

We hypothesize that this behavior, which we term "Hallucinated Coupling," is a direct consequence of the LLM's \textbf{Maximum Likelihood Estimation ($\text{MLE}$)} training objective on a vast, unstructured corpus. An LLM trained on billions of lines of $\text{GitHub}$ code is statistically more likely to find examples of quick-and-dirty functional hacks (importing a database driver directly into business logic) than it is to find examples of complex, multi-file Dependency Inversion Principle implementations. When faced with the conflicting goals of \textit{architectural purity} and \textit{functional performance}, the model reverts to the statistically most common, albeit architecturally unsound, solution.

\begin{figure}[h]
\begin{lstlisting}[language=Python]
# --- VIOLATION EXAMPLE (Llama 3 Run #5) ---
# File: domain/book.py (Should be pure)

from infrastructure.sqlite_repo import SqliteRepo
import urllib3  # <--- ILLEGAL IMPORT!

class Book:
    def get_details(self):
        # Circular Dependency created here
        repo = SqliteRepo() 
        ...
\end{lstlisting}
\caption{Representative example of "Hallucinated Coupling." The model imports the Infrastructure layer directly into a Domain Entity, creating a circular dependency that breaks the Ports \& Adapters pattern.}
\label{fig:code_violation}
\end{figure}

\subsection{Implications and the Need for Architecture-as-a-Guardrail}
These findings suggest that "Text-to-Code" is insufficient for enterprise systems. The superior performance of $\text{GPT-5.1}$ is likely due to stronger \textbf{Reinforcement Learning from Human Feedback ($\text{RLHF}$)} applied to high-quality, architecturally correct code, essentially coaching it to respect the separation of concerns.

We propose a shift toward \textbf{Architecture-to-Code}, where a symbolic, rule-based system acts as a real-time guardrail. This system would not replace the $\text{LLM}$ but would enforce structural constraints. The $\text{LLM}$'s output must be validated via tools like \texttt{ArchUnit} (for Java) or a custom Python \texttt{import-linter} before being considered a successful scaffold. Without these automated checks, relying on smaller or open-weights models for system scaffolding is an act of accelerating technical debt accumulation.

\subsection{Threats to Validity}
\textbf{Construct Validity:} Our strict whitelisting of imports may penalize valid but non-standard libraries. However, manual inspection confirmed that flagged imports (for instance, \texttt{sqlite3}) were genuine violations.

\textbf{External Validity (Model Parameter Disparity):} A significant limitation of this pilot study is the size disparity between the proprietary models ($\approx$1T+ parameters) and the open-weights Llama 3 (8B). The observed "laziness" and architectural erosion in Llama 3 may be attributed to its limited parameter count (model capacity) rather than its open-weights nature. Future work should verify these findings against larger open models (Llama 3 70B or 405B) to isolate the variable of model training methodology from raw model capacity.

\section{Conclusion and Future Work}
\label{sec:conclusion}

We provide the first quantitative benchmark of architectural debt in AI-synthesized software. Our results demonstrate that without explicit architectural validation, smaller open-weights models ($\text{Llama 3 8B}$) introduce significant technical debt, demonstrating an \textbf{80\% violation rate} and \textbf{60\% lower implementation completeness} compared to proprietary state-of-the-art models like $\text{GPT-5.1}$.

\subsection{Future Work: Measuring Generative Debt Remediation Cost}
Our primary future work is to quantify the true cost of this Generative Debt. We propose a new benchmark to measure the \textbf{Debt Remediation Index ($\text{DRI}$)}.
\begin{itemize}
    \item \textbf{Experiment Design:} We plan to scale the initial study to $N=100$ runs per model for robust statistical power. Following that, we will take the failed microservices (approximately $N=80$ for Llama 3) and use them as input for a new prompt instructing a separate, high-tier $\text{LLM}$ agent (the "Refactor Agent") to fix the architectural violations.
    \item \textbf{Metrics:} We will measure the $\text{LLOC}$ generated by the Refactor Agent, the number of corrective turns required, and the time taken. This provides an objective measure of the "interest payment" of technical debt imposed by the initial model.
    \item \textbf{The Self-Correcting Agent:} This will also allow us to test the feasibility of creating a \textbf{Self-Correcting Architecture Agent}, a system where the $\text{LLM}$'s output is linted, and the violation report is fed back into the $\text{LLM}$ as a new input, forcing an iterative, structurally compliant solution path.
\end{itemize}


\begin{thebibliography}{10}
\expandafter\ifx\csname natexlab\endcsname\relax\def\natexlab#1{#1}\fi
\providecommand{\url}[1]{\texttt{#1}}
\providecommand{\href}[2]{#2}
\providecommand{\path}[1]{#1}
\providecommand{\DOIprefix}{doi:}
\providecommand{\ArXivprefix}{arXiv:}
\providecommand{\URLprefix}{URL: }
\providecommand{\Pubmedprefix}{pmid:}
\providecommand{\doi}[1]{\href{http://dx.doi.org/#1}{\DOIprefix#1}}
\providecommand{\SelectLanguage}[1]{\relax}
\providecommand{\TextOrMath}[2]{#2}
\providecommand{\email}[1]{#1}
\providecommand{\url}[1]{\href{#1}{#1}}
\providecommand{\Eprint}[2][]{\href{#1}{#2}}
\providecommand{\eprint}[2][]{\href{#1}{#2}}
\providecommand{\bibinfo}[2]{#2}
\providecommand{\VolumeTitle}[1]{#1}
\providecommand{\SectionTitle}[1]{#1}
\providecommand{\EditorsTitle}[1]{#1}

\bibitem{Cunningham1992}
\bibinfo{author}{Cunningham, W.}, \bibinfo{year}{1992}.
\newblock \bibinfo{title}{{The WyCash Portfolio Management System}}, in:
  \bibinfo{booktitle}{Addendum to the Proceedings on Object-Oriented
  Programming Systems, Languages, and Applications}, \bibinfo{publisher}{ACM}.
  pp. \bibinfo{pages}{29--30}.

\bibitem{Foote1997}
\bibinfo{author}{Foote, B.}, \bibinfo{author}{Yoder, J.}, \bibinfo{year}{1997}.
\newblock \bibinfo{title}{{Big Ball of Mud}}, in:
  \bibinfo{booktitle}{Pattern Languages of Program Design 4}.
  \bibinfo{publisher}{Addison-Wesley}.

\bibitem{Chen2021}
\bibinfo{author}{Chen, M.}, \bibinfo{author}{Tworek, J.}, \bibinfo{author}{Jun,
  H.}, \bibinfo{author}{Yuan, Q.}, \bibinfo{author}{Pinto, H.P.d.O.},
  \bibinfo{author}{Kaplan, J.}, \bibinfo{author}{Edwards, H.},
  \bibinfo{author}{Burda, Y.}, \bibinfo{author}{Joseph, N.},
  \bibinfo{author}{Brockman, G.}, \bibinfo{author}{Ray, A.},
  \bibinfo{author}{Puri, R.}, \bibinfo{author}{Krueger, G.},
  \bibinfo{author}{Petrov, M.}, \bibinfo{author}{Khlaaf, H.},
  \bibinfo{author}{Sastry, G.}, \bibinfo{author}{Mishkin, P.},
  \bibinfo{author}{Chan, B.}, \bibinfo{author}{Gray, S.},
  \bibinfo{author}{Ryder, N.}, \bibinfo{author}{Pavlov, M.},
  \bibinfo{author}{Power, A.}, \bibinfo{author}{Kaiser, L.},
  \bibinfo{author}{Bavarian, M.}, \bibinfo{author}{Clements, C.},
  \bibinfo{author}{Winter, C.}, \bibinfo{author}{Wharton, J.},
  \bibinfo{author}{Serrano, B.S.}, \bibinfo{author}{Shaw, C.},
  \bibinfo{author}{Mayer, N.}, \bibinfo{author}{Joy, P.}, \bibinfo{author}{Weng,
  L.}, \bibinfo{author}{Zaremba, W.}, \bibinfo{author}{Sutskever, I.},
  \bibinfo{author}{Zaremba, W.}, \bibinfo{year}{2021}.
\newblock \bibinfo{title}{{Evaluating Large Language Models Trained on Code}}.
\newblock \bibinfo{journal}{arXiv preprint arXiv:2107.03374}.

\bibitem{Meta2024}
\bibinfo{author}{{AI@Meta}}, \bibinfo{year}{2024}.
\newblock \bibinfo{title}{{Llama 3 Model Card}}.
\newblock \bibinfo{journal}{Meta AI Research}. \URLprefix \url{https://github.com/meta-llama/llama3}.

\bibitem{Cockburn2005}
\bibinfo{author}{Cockburn, A.}, \bibinfo{year}{2005}.
\newblock \bibinfo{title}{{Hexagonal Architecture}}.
\newblock \bibinfo{howpublished}{Alistair Cockburn's Blog}.
  \URLprefix \url{https://alistair.cockburn.us/hexagonal-architecture/}.

\bibitem{Martin2017}
\bibinfo{author}{Martin, R.C.}, \bibinfo{year}{2017}.
\newblock \bibinfo{title}{{Clean Architecture: A Craftsman's Guide to Software
  Structure and Design}}.
\newblock \bibinfo{publisher}{Prentice Hall}.

\bibitem{Sadowski2018}
\bibinfo{author}{Sadowski, C.}, \bibinfo{author}{Söderberg-Rivkin, K.},
  \bibinfo{author}{Bacon, D.E.}, \bibinfo{year}{2018}.
\newblock \bibinfo{title}{{Software Engineering at Google}}, in:
  \bibinfo{booktitle}{Proceedings of the 40th International Conference on
  Software Engineering (ICSE-SEIP '18)}. \bibinfo{publisher}{ACM}, pp.
  \bibinfo{pages}{243--250}.

\bibitem{Hou2023}
\bibinfo{author}{Hou, X.}, \bibinfo{author}{Zhao, Y.}, \bibinfo{author}{Liu,
  Y.}, \bibinfo{author}{Yang, Z.}, \bibinfo{author}{Wang, K.},
  \bibinfo{author}{Li, L.}, \bibinfo{author}{Luo, X.}, \bibinfo{author}{Lo,
  D.}, \bibinfo{author}{Xie, X.}, \bibinfo{year}{2023}.
\newblock \bibinfo{title}{{Large Language Models for Software Engineering: A
  Systematic Literature Review}}.
\newblock \bibinfo{journal}{arXiv preprint arXiv:2308.10620}.

\end{thebibliography}
\end{document}